\documentclass[prl,twocolumn,groupedaddress,showpacs,nofootinbib]{revtex4}
\usepackage{graphicx}
\usepackage{dcolumn}
\usepackage{amssymb}
\usepackage{mathrsfs}
\usepackage{amsmath}
\usepackage{epsfig}
\usepackage[dvips]{color}
\usepackage{hhline}

\begin{document}

\title{WIMP Dark Matter and Baryogenesis}

\author{Pei-Hong Gu$^{1}_{}$}
\email{peihong.gu@mpi-hd.mpg.de}

\author{Manfred Lindner$^{1}_{}$}
\email{manfred.lindner@mpi-hd.mpg.de}

\author{Utpal Sarkar$^{2}_{}$}
\email{utpal@prl.res.in}

\author{Xinmin Zhang$^{3}_{}$}
\email{xmzhang@ihep.ac.cn}

\affiliation{$^{1}_{}$Max-Planck-Institut f\"{u}r Kernphysik,
Saupfercheckweg
1, 69117 Heidelberg, Germany\\
$^{2}_{}$Physical Research Laboratory, Ahmedabad 380009, India\\
$^{3}_{}$Institute of High Energy Physics, Chinese Academy of
Sciences, Beijing 100049, China}

\begin{abstract}

In the present universe visible and dark matter contribute
comparable energy density although they have different properties.
This coincidence can be elegantly explained if the dark matter relic
density, originating from a dark matter asymmetry, is fully
determined by the baryon asymmetry. Thus the dark matter mass is not
arbitrary, rather becomes predictive. We realize this scenario in
baryon(lepton) number conserving models where two or more neutral
singlet scalars decay into two or three baryonic(leptonic) dark
matter scalars, and also decay into quarks(leptons) through other
on-shell and/or off-shell exotic scalar bilinears. The produced
baryon(lepton) asymmetries in the dark matter scalar and in the
standard model quarks(leptons) are thus equal and opposite. The dark
matter mass can be predicted in a range from a few GeV to a few TeV
depending on the baryon(lepton) numbers of the decaying scalars and
the dark matter scalar. The dark matter scalar can interact with the
visible matter through the exchange of the standard model Higgs
boson, opening a window for the dark matter direct detection
experiments. These models also provide testable predictions in the
searches for the exotic scalar bilinears at LHC.

\end{abstract}


\maketitle

\emph{Introduction}: Cosmological observations indicate that dark
and visible matter have different properties but contribute
comparable energy density to the present universe
\cite{dunkley2008}. This intriguing coincidence indicates a common
origin of the dark and visible matter. The visible matter exists in
the present universe due to a matter--antimatter asymmetry, which is
the same as the baryon asymmetry. The dark matter relic density may
also be an asymmetry between the dark matter and antimatter
\cite{kuzmin1997,ki2005}. If the dark matter asymmetry is further
determined by the baryon asymmetry, we can naturally explain the
coincidence between the visible and dark matter. In this case the
dark matter mass is not arbitrary but predictive and depends on the
well known nucleon mass. In the existing models of the dark matter
asymmetry, the dark matter mass is expected to be in the GeV scale
unless we abandon the requirement of determining the baryon
asymmetry in terms of the dark matter asymmetry.

In this paper we propose a novel idea to predict the dark matter
mass in a range from the GeV scale to the TeV scale although the
dark matter asymmetry is fully determined by the baryon asymmetry.
In our models, the baryon(lepton) asymmetry in a dark matter scalar
\cite{sz1985} is equal but opposite to the baryon(lepton) asymmetry
in the standard model (SM) quarks(leptons). Although the total
baryon(lepton) asymmetry is zero, only the baryon(lepton) asymmetry
in the SM quarks(leptons) can be partially converted to the final
baryon asymmetry because the baryon(lepton) asymmetry in the dark
matter scalar has no effect on the sphaleron processes
\cite{krs1985}. As for the dark matter asymmetry, it is the ratio of
the baryon(lepton) asymmetry in the dark matter scalar over the
baryon(lepton) number of the dark matter scalar. We thus can expect
a smaller dark matter asymmetry and a heavier dark matter mass as
long as the dark matter scalar has a proper baryon(lepton) number.
This can be achieved if the decaying scalars directly decay into two
or three dark matter scalars while their decays into a number of the
SM quarks(leptons) are mediated by some on-shell and/or off-shell
exotic scalar bilinears, which couple to two SM quarks(leptons). Our
dark matter scalar is a typical weakly interacting massive particle
(WIMP) as it interacts with the visible matter through the exchange
of the SM Higgs boson.

When we were finalizing our manuscript, Buckley and Randall
\cite{br2010} have discussed the same idea to predict a bigger dark
matter mass from a smaller dark matter asymmetry. However, our
simple mechanism for suppressing the dark matter asymmetry is
different from theirs.

\emph{The Models}: We first construct the baryon number conserving
models. We introduce two color-triplet and iso-singlet scalars
$\delta(\textbf{3},\textbf{1},-\frac{1}{3})$ and
$\omega(\textbf{3},\textbf{1},\frac{2}{3})$, both of which carry a
baryon number $B=-\frac{2}{3}$. Their allowed Yukawa couplings to
the SM quark doublets $q_L^{}(\textbf{3},\textbf{2},\frac{1}{6})$
and singlets $u_R^{}(\textbf{3},\textbf{1},\frac{2}{3})$ and
$d_R^{}(\textbf{3},\textbf{1},-\frac{1}{3})$ are given by
\begin{eqnarray}
\label{lagrangian11} \mathcal{L}\supset
-f_\delta^{}\delta\bar{q}_{L}^c i\tau_2^{}q_L^{}-f'^{}_\delta\delta
\bar{u}_R^c d_R^{}-f_\omega^{} \omega \bar{d}_R^c d_R^{}
+\textrm{H.c.}\,.
\end{eqnarray}
We also introduce a neutral singlet scalar
$X_1^{}(\textbf{1},\textbf{1},0)$ with a baryon number $B=2$. So,
the following scalar interaction is allowed as it conserves the
baryon number,
\begin{eqnarray}
\label{lagrangian12} \mathcal{L}\supset -\kappa_1^{} X_1^{} \omega
\delta^2_{} +\textrm{H.c.}\,.
\end{eqnarray}
There can be more neutral singlet scalars
$X_i^{}(\textbf{1},\textbf{1},0)$ with the baryon numbers defined by
\begin{eqnarray}
\label{lagrangian13} \mathcal{L}\supset -\sum_{i=2}^{n}\kappa_i^{}
X_i^\ast X_{i-1}^{a_{i-1}^{}}
+\textrm{H.c.}~\textrm{with}~a_{i-1}^{}=2~\textrm{or}~3\,.
\end{eqnarray}
Finally we introduce another neutral singlet scalar
$\chi^{}_{}(\textbf{1},\textbf{1},0)$. In the case without
$X_i^{}(i=2,...,n)$, we consider the baryon number conserving
interaction between $\chi$ and $X_1^{}$,
\begin{eqnarray}
\label{lagrangian14} \mathcal{L}\supset -\gamma_1^{}
X_1^\ast\chi^b_{} +\textrm{H.c.}~\textrm{with}~b=2~\textrm{or}~3\,.
\end{eqnarray}
In the presence of $X_i^{}(i=2,...,n)$, we consider the baryon
number conserving interaction between $\chi$ and $X_n^{}$,
\begin{eqnarray}
\label{lagrangian15} \mathcal{L}\supset -\gamma_n^{}
X_n^\ast\chi^b_{}
+\textrm{H.c.}~\textrm{with}~b=2~\textrm{or}~3~\textrm{but}~b\neq
a_{n-1}^{} \,.
\end{eqnarray}
Here the choice $b\neq a_{n-1}^{}$ is to forbid the couplings $
\mathcal{L}\supset
-\alpha\chi^\ast_{}X_{n-1}^{}-\beta(\chi^\ast_{}X_{n-1}^{})^2_{}+\textrm{H.c.}$.
We emphasize that by proper choice of the parameters in the scalar
potential the neutral scalars $X_i^{}$ and $\chi$ will not develop
any vacuum expectation values (VEVs) to break the baryon number. So,
we can have a stable $\chi$ to act as the dark matter. In a similar
fashion we could consider color-sextet scalars to implement this
scenario.

We also construct models with conserving lepton number. We introduce
an iso-triplet scalar $\xi(\textbf{1},\textbf{2},1)$ with a lepton
number $L=-2$. The triplet scalar $\xi$ thus can have the Yukawa
couplings with the SM lepton doublets
$l_L^{}(\textbf{1},\textbf{2},-\frac{1}{2})$,
\begin{eqnarray}
\label{lagrangian21} \mathcal{L}\supset -f_\xi^{}\bar{l}_{L}^c
i\tau_2^{} \xi l_L^{} +\textrm{H.c.}\,.
\end{eqnarray}
We then introduce a neutral singlet
$X_1^{}(\textbf{1},\textbf{1},0)$ with a lepton number $L=2$ and
other neutral singlet scalars
$X_i^{}(\textbf{1},\textbf{1},0)(i=2,...,n)$ with the lepton numbers
defined by
\begin{eqnarray}
\label{lagrangian22} \mathcal{L}\supset -\sum_{i=2}^{n}\kappa_i^{}
X_i^\ast X_{i-1}^{a_{i-1}^{}}
+\textrm{H.c.}~\textrm{with}~a_{i-1}^{}=2~\textrm{or}~3\,.
\end{eqnarray}
The singlet scalar $X_1^{}$ has a lepton number conserving
interaction with the triplet scalar $\xi$ and the SM Higgs doublet
$\phi(\textbf{1},\textbf{2},-\frac{1}{2})$,
\begin{eqnarray}
\label{lagrangian25} \mathcal{L}\supset -\kappa_1^{} X_1^{}
\phi^T_{}i\tau_2^{}\xi \phi +\textrm{H.c.}\,.
\end{eqnarray}
We further introduce another neutral singlet scalar
$\chi^{}_{}(\textbf{1},\textbf{1},0)$ with the lepton number defined
by
\begin{eqnarray}
\label{lagrangian23} \mathcal{L}\supset -\gamma_1^{}
X_1^\ast\chi^b_{} +\textrm{H.c.}~\textrm{with}~b=2~\textrm{or}~3
\end{eqnarray}
in the absence of $X_i^{}(i=2,...,n)$, or by
\begin{eqnarray}
\label{lagrangian24} \mathcal{L}\supset -\gamma_n^{}
X_n^\ast\chi^b_{}
+\textrm{H.c.}~\textrm{with}~b=2~\textrm{or}~3~\textrm{but}~b\neq
a_{n-1}^{} \,.
\end{eqnarray}
in the presence of $X_i^{}(i=2,...,n)$. The singlet $\chi$ can keep
stable to be the dark matter as a result of the unbroken lepton
number. We can replace the triplet scalar $\chi$ by two charged
singlet scalars $\zeta(\textbf{1},\textbf{1},1)$ and
$\varrho(\textbf{1},\textbf{1},\textbf{2})$, both of which carry a
lepton number $L=-2$, to construct the lepton number conserving
models. Specifically, the interactions (\ref{lagrangian21}) and
(\ref{lagrangian25}) should be replaced by
\begin{eqnarray}
\label{lagrangian26} \mathcal{L}\supset -f_\zeta^{}\zeta
\bar{l}_{L}^c i\tau_2^{} l_L^{}-f_\varrho^{}\varrho \bar{e}_R^c
e_R^{}-\kappa_1^{} X_1^{}\varrho^\ast_{}\zeta^2 +\textrm{H.c.}\,,
\end{eqnarray}
with $e_R^{}(\textbf{1},\textbf{1},-1)$ being the right-handed
charged leptons.

\begin{figure}
\vspace{3.2cm} \epsfig{file=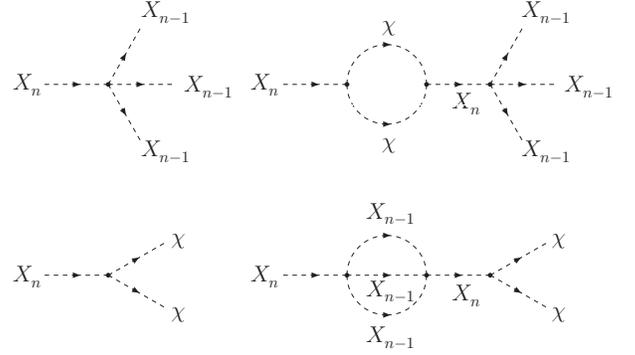, bbllx=5.3cm, bblly=6.0cm,
bburx=15.3cm, bbury=16cm, width=6.cm, height=6.cm, angle=0, clip=0}
\vspace{-4.5cm} \caption{\label{baryogenesis} An example of the
decays of the singlet scalars $X_n^{}$ for generating the
baryon(lepton) asymmetries in the subsequently decaying scalar
$X_{n-1}^{}$ and in the dark matter scalar $\chi$. The subsequent
decays into the SM quarks(leptons) are not shown for simplicity.}
\end{figure}

\emph{Dark Matter Asymmetry and Mass}: In the above baryon(lepton)
number conserving models, the neutral singlet scalars
$X_k^{}(k=1~\textrm{or}~n)$ have two decay channels: one is into two
or three dark matter scalars $\chi$, the other one is into a number
of the SM quarks(leptons) through other on-shell and/or off-shell
scalars. In the presence of two or more $X_k^{}
(k=1~\textrm{or}~n)$, we can obtain a baryon(lepton) asymmetry in
the SM quarks(leptons) and an equal but opposite baryon asymmetry in
the dark matter scalar $\chi$,
\begin{eqnarray}
\varepsilon_\chi^{B(L)}=-\varepsilon_\textrm{SM}^{B(L)}\,.
\end{eqnarray}
This can be understood from Fig. \ref{baryogenesis} where we show an
example of the decays of $X_n^{}$. Since the baryon(lepton)
asymmetry in the dark matter scalar has no effect on the sphaleron
processes, the baryon(lepton) asymmetry in the SM quarks(leptons)
can be partially converted to the final baryon asymmetry
\cite{krs1985},
\begin{eqnarray}
\eta_B^{}= \frac{28}{79} \varepsilon_\textrm{SM}^{B} ~~\textrm{or}~~
\eta_B^{}=- \frac{28}{79} \varepsilon_\textrm{SM}^{L}\,.
\end{eqnarray}
This is like the essence of the leptogenesis \cite{fy1986} with
Dirac neutrinos \cite{dlrw2000}. Here the other interactions that
violate the SM baryon and/or lepton number have been assumed to
decouple before the above baryogenesis epoch. We also have assumed
that the baryon(lepton) asymmetry is produced before the electroweak
phase transition. This assumption is not necessary for the baryon
number conserving models, i.e. the coefficient $\frac{28}{79}$
should be absent if the baryon asymmetry is produced after the
electroweak phase transition. We shall not consider this possibility
for simplicity.

As for the dark matter asymmetry, it is the ratio of the
baryon(lepton) asymmetry in the dark matter scalar over the
baryon(lepton) number of the dark matter scalar,
\begin{eqnarray}
\eta_\chi^{}=\left\{\begin{array}{lcl} \pm \frac{b}{2}
\varepsilon_\chi^{B(L)} & \textrm{for} & k=1\,,\\
[2mm] \pm
\frac{b}{2a_1^{}a_2^{}...a_{n-1}^{}}\varepsilon_\chi^{B(L)}&\textrm{for}&k=n\,.\end{array}
\right.
\end{eqnarray}
Here the choice of the sign $+$ or $-$ depends on the definition of
the dark matter and dark antimatter, i.e. $\chi$ is the dark matter or
the dark antimatter. If the dark matter asymmetry is responsible for
the dark matter relic density, we can read
\begin{eqnarray}
\Omega_B^{}: \Omega_\chi^{}=\eta_B^{} m_N^{}:\eta_\chi^{} m_\chi^{}
\,,
\end{eqnarray}
where $m_N^{}$ is the nucleon mass and will be taken to be
$m_N^{}=\frac{1}{2}(m_p^{}+m_n^{})= 939\,\textrm{MeV}$ by ignoring
the tiny difference between the proton and neutron masses. Since the
fractions of the visible and dark matter in the present universe has
been precisely measured, i.e. \cite{dunkley2008}
\begin{eqnarray}
\Omega_B^{}h^2_{}&=&0.02273\pm0.00062\,,\nonumber\\
\Omega_\chi^{}h^2_{}&=&0.1099\pm0.0062\,,
\end{eqnarray}
we can predict the dark matter mass by
\begin{eqnarray}
\label{dmmass}
m_\chi^{}=\frac{\Omega_\chi^{}}{\Omega_B^{}}\frac{\eta_B^{}}{\eta_\chi^{}}m_N^{}\,.
\end{eqnarray}

The above predictive dark matter mass can be in a range from the GeV
scale to the TeV scale by choosing the baryon(lepton) number of the
decaying neutral scalars and the dark matter scalar. For example, in
the case with $k=1$, we read
\begin{eqnarray}
\label{dmmass1} m_\chi^{}=\left\{\begin{array}{rcc}
1.07\,\textrm{GeV}&\textrm{for} & b=3\,,\\
[2mm]  1.61\,\textrm{GeV}&\textrm{for} & b=2\,.\end{array}\right.
\end{eqnarray}
It is possible to predict a heavier dark matter mass by choosing a
bigger baryon(lepton) number of the dark matter scalar. For example,
by taking $a_1^{}=...=a_{n-1}^{}=3$ and $b=2$, we read
\begin{eqnarray}
\label{dmmass2} m_\chi^{}=\left\{\begin{array}{rcc}
4.83\,\textrm{GeV} & \textrm{for} & n=2\,,\\
[2mm] 14.5\,\textrm{GeV}& \textrm{for} & n=3\,,\\
[2mm] 43.4\,\textrm{GeV}& \textrm{for} & n=4\,,\\
[2mm] 130\,\textrm{GeV}& \textrm{for} & n=5\,,\\
[2mm] 391\,\textrm{GeV}& \textrm{for} & n=6\,,\\
[2mm] 1.17\,\textrm{TeV}& \textrm{for} & n=7\,.
\end{array}\right.
\end{eqnarray}

\emph{Dark Matter Direct Detection}: The dark matter scalar $\chi$
has a quartic coupling with the SM Higgs doublet $\phi$,
\begin{eqnarray}
\mathcal{L}\supset-\lambda_{\chi\phi}^{}\chi^\ast_{} \chi
\phi^\dagger_{}\phi\,,
\end{eqnarray}
and hence a trilinear coupling with the SM Higgs boson,
\begin{eqnarray}
\mathcal{L}\supset-\lambda_{\chi\phi}^{}vh\chi^\ast_{} \chi
~~\textrm{with}~~\phi = \left[
\begin{array}{c}
\frac{1}{\sqrt{2}}(v+h)\\
[2mm] 0 \end{array}\right]\,.
\end{eqnarray}
The t-channel exchange of the SM Higgs boson $h$ will result in an
elastic scattering of the dark matter $\chi$ on the nucleon $N$. The
dark matter scattering cross section would be
\begin{eqnarray}
\label{crosssectiondn} \sigma_{\chi N\rightarrow \chi N }^{}
&=&\frac{\lambda_{\chi\phi}^2}{4\pi}\frac{f^2_{}m_N^2 \mu_r^2
}{m_h^4 m_\chi^2} \,,
\end{eqnarray}
where $\mu_{r}^{}=m_\chi^{} m_N^{}/(m_\chi^{} + m_N^{})$ is the
reduced mass, the factor $f$ in the range $0.14 <f<0.66$ has a
central value $f=0.30$ \cite{aht2008}. The scattering cross section
(\ref{crosssectiondn}) is strongly constrained by the dark matter
direct detection experiments \cite{ksz2009}. We find for the
predictive dark matter masses (\ref{dmmass1}) and (\ref{dmmass2}),
the dark matter scattering cross section can fulfill the
experimental results for an appropriate $\lambda_{\chi\phi}^{}$. For
example, by fixing $m_h^{}=120\,\textrm{GeV}$ and $f=0.3$, we read
\begin{eqnarray}
\sigma_{\chi N\rightarrow \chi N }^{}= \left\{\begin{array}{c}
\frac{\lambda_{\chi\phi}^2}{1}\cdot 2.6\cdot
10^{-39}_{}\,\textrm{cm}^2_{}~\textrm{for}~m_\chi=1.07\,\textrm{GeV}\,,\\
[2mm] \frac{\lambda_{\chi\phi}^2}{1}\cdot 1.6\cdot
10^{-39}_{}\,\textrm{cm}^2_{}~\textrm{for}~m_\chi=1.61\,\textrm{GeV}\,,\\
[2mm] \frac{\lambda_{\chi\phi}^2}{1}\cdot 3.1\cdot
10^{-40}_{}\,\textrm{cm}^2_{}~\textrm{for}~m_\chi=4.83\,\textrm{GeV}\,,\\
[2mm] \frac{\lambda_{\chi\phi}^2}{0.1}\cdot 4.4\cdot
10^{-42}_{}\,\textrm{cm}^2_{}~\textrm{for}~m_\chi=14.5\,\textrm{GeV}\,,\\
[2mm] \frac{\lambda_{\chi\phi}^2}{0.01}\cdot 5.3\cdot
10^{-44}_{}\,\textrm{cm}^2_{}~\textrm{for}~m_\chi=43.4\,\textrm{GeV}\,,\\
[2mm] \frac{\lambda_{\chi\phi}^2}{0.05}\cdot 3.0\cdot
10^{-44}_{}\,\textrm{cm}^2_{}~\textrm{for}~m_\chi=130\,\textrm{GeV}\,,\\
[2mm] \frac{\lambda_{\chi\phi}^2}{1}\cdot 6.8\cdot
10^{-44}_{}\,\textrm{cm}^2_{}~\textrm{for}~m_\chi=391\,\textrm{GeV}\,,\\
[2mm] \frac{\lambda_{\chi\phi}^2}{5}\cdot 3.8\cdot
10^{-44}_{}\,\textrm{cm}^2_{}~\textrm{for}~m_\chi=1.17\,\textrm{TeV}\,.\end{array}\right.
\end{eqnarray}

\emph{Dark Matter and Antimatter Annihilation}: If the dark matter
asymmetry is expected to account for the dark matter relic density,
a fast annihilation between the dark matter and antimatter should be
guaranteed to dilute the thermally produced relic density. In other
words, the annihilation cross section should be much bigger than the
typical value $\sim 1\,\textrm{pb}$ for thermally generating the
dark matter relic density. We check for a dark matter mass above a
few hundred GeV, the dark matter scalar can annihilate very fast
into the SM fields through its quartic coupling with the SM Higgs
doublet. For example, we have
\begin{eqnarray}
\langle\sigma v\rangle =\frac{\lambda_{\chi\phi}^2}{16\pi
m_\chi^2}=\left\{\begin{array}{rcl}
 \frac{\lambda_{\chi\phi}^2}{1}\cdot 50.7\,\textrm{pb}&\textrm{for}&m_\chi^{}=~391\,\textrm{GeV}\,,\\
[2mm] \frac{\lambda_{\chi\phi}^2}{5}\cdot
28.3\,\textrm{pb}&\textrm{for}&m_\chi^{}=1.17\,\textrm{TeV}\,.
\end{array}\right.
\end{eqnarray}

As for the case with a dark matter mass around or below the SM Higgs
mass, the dark matter annihilation into the SM fields should be
suppressed by the quartic coupling $\lambda_{\chi^{}\phi}^{}$
(constrained by the dark matter direct detection experiments) and/or
the Yukawa couplings of the SM fermions. In this case, we need to
introduce other fields to enhance the annihilation between the dark
matter and antimatter. For example, we can consider a Higgs singlet
$\sigma(\textbf{1},\textbf{1},0)$ to break a global symmetry at the
electroweak scale, i.e.
\begin{eqnarray}
\sigma=\frac{1}{\sqrt{2}}(v'+h')e^{i\frac{\rho}{v'}}_{} \,.
\end{eqnarray}
The dark matter scalar $\chi$ has a quartic coupling with $\sigma$,
\begin{eqnarray}
\mathcal{L}&\supset &-\lambda_{\chi\sigma}^{}\chi^\ast_{}\chi
\sigma^\ast_{}\sigma \,,
\end{eqnarray}
so that it can significantly annihilate into the massless Goldstone
$\rho$. For example, by fixing $m_{h'}^{}=70\,\textrm{GeV}$, we have
\begin{eqnarray}
\langle\sigma v\rangle =\frac{\lambda_{\chi\sigma}^2 m_\chi^2}{4\pi
m_{h'}^4}=\left\{\begin{array}{rcl}
 \frac{\lambda_{\chi\sigma}^2}{10}\cdot 14.8\,\textrm{pb}&\textrm{for}&m_\chi^{}=1.07\,\textrm{GeV}\,,\\
[2mm] \frac{\lambda_{\chi\phi}^2}{5}\cdot
16.7\,\textrm{pb}&\textrm{for}&m_\chi^{}=1.61\,\textrm{GeV}\\
[2mm] \frac{\lambda_{\chi\phi}^2}{1}\cdot
30.1\,\textrm{pb}&\textrm{for}&m_\chi^{}=4.83\,\textrm{GeV}\,.
\end{array}\right.
\end{eqnarray}
Here we simply ignored the mixing between the SM Higgs boson
$h$ and the non-SM one $h'$ which is from the quartic coupling,
\begin{eqnarray}
\mathcal{L}&\supset &-\lambda_{\sigma\phi}^{} \sigma^\ast_{}\sigma
\phi^\dagger_{}\phi \,.
\end{eqnarray}
The $h-h'$ mixing will not significantly change the annihilation
cross section but will have interesting implications on the SM Higgs
searches at the colliders \cite{jl1991}. The global symmetry
breaking can be related to the Dirac \cite{rw1983,gh2006} seesaw
\cite{minkowski1977} mechanism for generating the small Dirac
neutrino masses. For example, we consider \cite{gh2006}
\begin{eqnarray}
\mathcal{L}&\supset &-y_\nu^{}\bar{l}_{L}^{}\eta
\nu_{R}^{}-\mu\sigma \eta^\dagger_{}\phi +\textrm{H.c.}\,,
\end{eqnarray}
where $\nu_R^{}(\textbf{1},\textbf{1},0)$ denotes the right-handed
neutrinos and $\eta(\textbf{1},\textbf{2},-\frac{1}{2})$ is a new
Higgs doublet. The global symmetry is imposed in the way that
$\nu_R^c$, $\eta$ and $\sigma$ carry the same quantum numbers. For
$\langle\sigma\rangle\sim \langle\phi\rangle$, we can obtain the
desired neutrino masses for $y_\nu^{}=\mathcal{O}(1)$ and
$\mu\lesssim m_\eta^{}=\mathcal{O}(10^{14}_{}\,\textrm{GeV})$. Note
the contribution from $\eta$ to the Goldstone $\rho$ should be
negligible since $\langle\eta\rangle\ll\langle\sigma\rangle$.

\emph{Summary}: In this paper we proposed a novel mechanism to
predict the dark matter mass in the range from the GeV scale to the
TeV scale. In our scenario, the dark matter relic density is a
nonthermally \cite{lhzb2000} produced dark matter asymmetry
determined by the baryon asymmetry so that we can naturally explain
the comparable energy density of the visible and dark matter in the
present universe. The dark matter mass is thus predictive rather
than arbitrary. In particular, the predictive dark matter mass
should depend on the baryon(lepton) number of the dark matter field
since the dark matter asymmetry is the ratio of the baryon(lepton)
asymmetry in the dark matter field over the baryon(lepton) number of
the dark matter field. This means that the dark matter can have a
heavier mass if it has a bigger baryon(lepton) number. We
demonstrated this possibility by constructing some renormalizable
models, where the decaying scalars directly decay into two or three
dark matter scalars while their decays into a number of the SM
quarks(leptons) are mediated by some on-shell and/or off-shell
scalar bilinears. Our dark matter scalar is consistent with the
present dark matter direct detection experiments and can be verified
in the future. The required scalar bilinears can also be detected at
LHC or ILC. In our examples, we have not discussed explicitly the
generation of the decaying scalars, which could be thermally
produced by their interactions with other fields. This is the
thermal baryogenesis scenario. Alternatively, we can consider the
nonthermal baryogenesis, where the decaying scalars are responsible
for the chaotic inflation \cite{linde1983}.

\textbf{Acknowledgement}: ML is supported by the
Sonderforschungsbereich TR 27 of the Deutsche
Forschungsgemeinschaft. PHG is supported by the Alexander von
Humboldt Foundation. XZ is supported by the National Natural Science
Foundation of China.


\begin{thebibliography}{99}


\bibitem{dunkley2008}
J. Dunkley {\it et al.}, [WMAP Collaboration], Astrophys. J. Suppl.
\textbf{180}, 306 (2009).

\bibitem{kuzmin1997}
V.A. Kuzmin, Phys. Part. Nucl. \textbf{29}, 257 (1998), Fiz. Elem.
Chast. Atom. Yadra \textbf{29}, 637 (1998); Phys. Atom. Nucl.
\textbf{61}, 1107 (1998).

\bibitem{ki2005}
R. Kitano and I. Low, Phys. Rev. D \textbf{71}, 023510 (2005); N.
Cosme, L. Lopez Honorez, and M.H.G. Tytgat, Phys. Rev. D
\textbf{72}, 043505 (2005); D.E. Kaplan, M.A. Luty, and K.M. Zurek,
Phys. Rev. D \textbf{79}, 115016 (2009); P.H. Gu, U. Sarkar, and X.
Zhang, Phys. Rev. D \textbf{80}, 076003 (2009); H. An, S.L. Chen,
R.N. Mohapatra, and Y. Zhang, JHEP \textbf{1003}, 124 (2010); P. Gu
and U. Sarkar, Phys. Rev. D \textbf{81}, 033001 (2010).



\bibitem{sz1985}
V. Silveira and A. Zee, Phys. Lett. B \textbf{161}, 136 (1985); J.
McDonald, Phys. Rev. D \textbf{50}, 3637 (1994); C.P. Burgess, M.
Pospelov, and T. ter Veldhuis, Nucl. Phys. B \textbf{619}, 709
(2001).



\bibitem{krs1985}
V.A. Kuzmin, V.A. Rubakov, and M.E. Shaposhnikov, Phys. Lett. B
\textbf{155}, 36 (1985).


\bibitem{br2010}
M.R. Buckley and L. Randall, arXiv:1009.0270 [hep-ph].



\bibitem{fy1986}
M. Fukugita and T. Yanagida, Phys. Lett. B \textbf{174}, 45 (1986).


\bibitem{dlrw2000}
K. Dick, M. Lindner, M. Ratz, and D. Wright, Phys. Rev. Lett.
\textbf{84}, 4039 (2000).




\bibitem{aht2008}
S. Andreas, T. Hambye, and M.H.G. Tytgat, JCAP \textbf{0810}, 034
(2008); and references therein.


\bibitem{ksz2009}
J. Kopp, T. Schwetz, and J. Zupan, JCAP \textbf{1002}, 014 (2010);
and references therein.


\bibitem{jl1991}
G. Jungman and M.A. Luty, Nucl. Phys. B \textbf{361}, 24 (1991); A.
Dedes, T. Figy, S. Hoche, F. Krauss, and T.E.J. Underwood, JHEP
\textbf{0811}, 036 (2008).


\bibitem{rw1983}
M. Roncadelli and D. Wyler, Phys. Lett. B \textbf{133}, 325 (1983);
P. Roy and O. Shanker, Phys. Rev. Lett. \textbf{52}, 713 (1984).




\bibitem{gh2006}
P.H. Gu and H.J. He, JCAP \textbf{0612}, 010 (2006).


\bibitem{minkowski1977}
P. Minkowski, Phys. Lett. \textbf{67B}, 421 (1977); T. Yanagida, in
{\it Proc. of the Workshop on Unified Theory and the Baryon Number
of the Universe}, ed. O. Sawada and A. Sugamoto (KEK, Tsukuba,
1979), p. 95; M. Gell-Mann, P. Ramond, and R. Slansky, in {\it
Supergravity}, ed. F. van Nieuwenhuizen and D. Freedman (North
Holland, Amsterdam, 1979), p. 315; S.L. Glashow, in {\it Quarks and
Leptons}, ed. M. L\'evy {\it et al.} (Plenum, New York, 1980), p.
707; R.N. Mohapatra and G. Senjanovi\'c, Phys. Rev. Lett.
\textbf{44}, 912 (1980).


\bibitem{lhzb2000}
W.B. Lin, D.H. Huang, X. Zhang, and R.H. Brandenberger, Phys. Rev.
Lett. \textbf{86}, 954 (2000).


\bibitem{linde1983}
A.D. Linde, Phys. Lett. B \textbf{129}, 177 (1983).





\end{thebibliography}
\end{document}